\documentclass[ reprint, superscriptaddress,grouped address, amsmath,assume, aps,materials,]{revtex4-2}

\usepackage{graphicx}
\usepackage{dcolumn}
\usepackage{bm}
\usepackage{physics}
\usepackage{xcolor}
\usepackage[normalem]{ulem}
\usepackage{hyperref}
\usepackage{float}
\usepackage{atbegshi}
\usepackage{emptypage}

\usepackage{pdfpages}

\newcounter{pagecounter}
\AtBeginShipout{%
  \stepcounter{pagecounter}%
  \ifnum\value{pagecounter}=17
    \AtBeginShipoutDiscard 
  \fi
}

\makeatletter
\AtBeginDocument{\let\LS@rot\@undefined}
\makeatother

\hypersetup{
     colorlinks   = true,
     citecolor    = blue,
     linkcolor    = blue,
     urlcolor     = blue,
}

\newcommand{\RNum}[1]{\uppercase\expandafter{\romannumeral #1\relax}}

\newcommand{\Rnum}[1]{\lowercase\expandafter{\romannumeral #1\relax}}

\begin{document}

\preprint{APS/123-QED}

\title{Core structure of dislocations in ordered ferromagnetic FeCo}

\author{Aleksei~Egorov}
\affiliation{ ICAMS, Ruhr-Universit\"{a}t Bochum, Universit\"{a}tstr. 150, 44780 Bochum, Germany }
\affiliation{ Engineering and Technology Institute (ENTEG), Faculty of Science and Engineering, University of Groningen, Groningen 9747AG, The Netherlands }

\author{Antoine~Kraych}
\affiliation{ ICAMS, Ruhr-Universit\"{a}t Bochum, Universit\"{a}tstr. 150, 44780 Bochum, Germany }

\author{Matous~Mrovec}
\affiliation{ ICAMS, Ruhr-Universit\"{a}t Bochum, Universit\"{a}tstr. 150, 44780 Bochum, Germany }

\author{Ralf~Drautz}
\affiliation{ ICAMS, Ruhr-Universit\"{a}t Bochum, Universit\"{a}tstr. 150, 44780 Bochum, Germany }

\author{Thomas~Hammerschmidt}
\affiliation{ ICAMS, Ruhr-Universit\"{a}t Bochum, Universit\"{a}tstr. 150, 44780 Bochum, Germany }

\begin{abstract}
We elucidated the core structure of screw dislocations in ordered B2 FeCo using a recent magnetic bond-order potential (BOP)~[Egorov~$et~al.$, \href{https://journals.aps.org/prmaterials/abstract/10.1103/PhysRevMaterials.7.044403}{Phys.~Rev.~Mater. {\bf 7}, 044403~(2023)}]. We corroborated that dislocations in B2 FeCo exist in pairs separated by antiphase boundaries. The equilibrium separation is about 50~\AA, which demands large-scale atomistic simulations—inaccessible for density functional theory but attainable with BOP. We performed atomistic simulations of these separated dislocations with BOP and predicted that they reside in degenerate core structures. Also, dislocations induce changes in the local electronic structure and magnetic moments.
\end{abstract}

\maketitle

\section{\label{sec:Introduction}Introduction}

Iron-cobalt combines excellent magnetic properties and severe brittleness~\cite{SunDeevi05}. The latter can be related to how the atoms arrange in a compact region surrounding the dislocation line—the dislocation core~\cite{anderson2017theory}.
Although many experimental~\cite{STOLOFFDAVIESS1964, MarcinkowskiChessin1964, baker1994, GEORGE2002325}, theoretical~\cite{sadananda-marcinkowski.1973-feco-disl-theor, marcinkowski1974cross-slip}, and computational~\cite{DFT_GS_GHOLIZADEH2018, Li2021eam, Muralles2022meam, MURALLES2023101670} studies have examined the mechanical properties of ordered Fe-Co alloys, the dislocation core structure remains unknown. Observing cores experimentally proves complicated~\cite{sigle1999-core-experiment-1, mendis2006-core-experiment-2-and-Nye-tensor, groger2011-core-experiment}, and quantum mechanical calculations based on density functional theory (DFT)~\cite{PhysRev.136.B864-DFT-1, PhysRev.140.A1133DFT-2} face difficulties due to the large simulation cells required for modeling ordered alloys with dissociated dislocations~\cite{KoehlerSeitzSuperdislo}.

\begin{figure}[!bp]
\begin{center}
\includegraphics[width=0.499\textwidth]{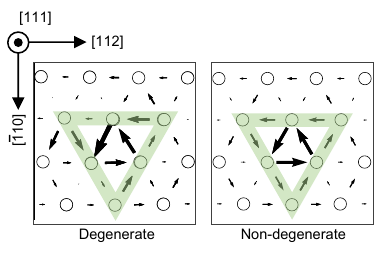}
\caption{Differential displacement illustrates the two possible core symmetries of the $\frac{1}{2}$[111] screw dislocations in bcc-like crystals—B2 FeCo (left) and bcc Fe (right) obtained with BOP~\cite{EgorovBOP23}. The arrows reveal the changes in atomic positions along the [111] direction of the dislocation line compared to a bulk defect-free cell. On the left, longer and shorter arrows alternating on a green triangle reveal the degenerate core. Conversely, on the right, the same length of the arrows reveals the non-degenerate core~\cite{DDP-comment}.
}
\label{fig:dislo_core_example}
\end{center}
\end{figure}

Dislocations carry plastic deformation~\cite{sutton2021concepts}. If a crystal deforms easily, dislocations must be mobile under low applied stress. When immobile, they cannot relieve stresses by shearing the crystal; cracks start to grow and material fractures~\cite{gordon2006new}. In metals with the body-centered cubic (bcc) structure, $\frac{1}{2}$[111] screw dislocations govern low-temperature plasticity due to their compact, non-planar core structure~\cite{duesberyvitek1973corebcc, christian1983some, vitek2008non, weygand2015multiscale}. Their core structures can be divided according to symmetry into degenerate and non-degenerate (see Fig.~\ref{fig:dislo_core_example}). DFT studies revealed that the non-degenerate core is the ground state for pure bcc metals ~\cite{Woodward_Mo_and_Ta_cores_DFT_2001, Ismail-Beigi_Mo_and_Ta_cores_DFT_2000, Frederiksen_cores_DFT_2003, Dezerald-dos-in-core, PhysRevB.87.054114-non-degenerate-core-bcc-TM}.
Degenerate cores, for a long time, remained just the artifacts produced by classical potentials~\cite{DudarEmpPotFailue}, until Romaner $et$~$al.$ observed degenerate cores with DFT in disordered bcc alloys, first in W-Re~\cite{Romaner_W-Re_PRL2010} and then in Fe-Co~\cite{Romaner_FeCo_2014}. The authors also speculated that the degenerate core exists in ordered B2 (CsCl) FeCo (the most stable phase at 1:1 composition~\cite{FeCoPhaDiaChap}) but exploring it with DFT was not feasible.

The behavior of screw dislocations in ordered and disordered Fe-Co alloys is distinctly different. In tensile tests, the ordered B2 FeCo fractures in a brittle manner without elongation, whereas the disordered FeCo before fracture displays some ductility~\cite{baker1994}. In the latter, wavy dislocation slip lines~\cite{MarcinkowskiChessin1964} indicate profuse cross-slip, similar to pure bcc transition metals. This similarity implies that in disordered FeCo, ordinary $\frac{1}{2}$[111] screw dislocations 
govern plasticity. In contrast, in (partially) ordered B2 FeCo, the straight slip lines indicate rare cross-slip. This behavior likely arises from disparate dislocations presented in B2 FeCo. When an ordinary $\frac{1}{2}$[111] screw dislocation glides in a B2 crystal, it disrupts the chemical order within the $\{110\}$ slip plane, producing an interface known as antiphase boundary (APB). 
Therefore, plasticity in B2 alloys is governed by so-called superdislocations~\cite{KoehlerSeitzSuperdislo, vitek2008non}, with Burgers vectors two times larger than those of ordinary ones. Superdislocation gliding through B2 crystal retains its chemical order.

In B2 FeCo, [111] screw superdislocations dissociate into two ordinary $\frac{1}{2}$[111] screw dislocations, called \textit{partial dislocations} or \textit{partials}, separated by an APB, which glide together (Fig.~\ref{fig:simcell})~\cite{MarcinkowskiChessin1964, moine1971-partials-pair-TEM-picture}. 
The equilibrium separation depends on the balance of two opposing forces: the elastic repulsion of the partials with the same Burgers vector counters the tension due to the energy cost of creating APB between them~\cite{marcinkowski1974-review}.
The estimated separation of the partials in B2 FeCo is about 50~Å~\cite{marcinkowski1963ExpAPBener}, and their modeling would entail simulation cells with thousands of atoms—unfeasible for DFT~\cite{Clouet2020-ab-initio}. 
Classical interatomic potentials, such as the embedded-atom method (EAM)~\cite{EAM} and the modified EAM (MEAM)~\cite{MEAM-1, MEAM-2, MEAM-3}, easily handle large simulation cells but only crudely describe the angular character of the bonding, vital for bcc transition metals~\cite{vitek2004-angular-dependence-2, NGUYENMANH2007255-angular-dependence}. This renders EAM and MEAM ambiguous for core structures~\cite{DudarEmpPotFailue}.
Besides, these potentials usually lack magnetism (and magnetism is the fulcrum of the Fe-Co properties~\cite{hawkins1988ferromagnetism-and-stability-fe-co, AbrikFeCoPRB, FahnleAntisite}). Machine learning~(ML) interatomic potentials combine exceptional accuracy and efficiency~\cite{deringer2019MLreview-1,mishin2021MLreview-2,behler2021review-4,kulik2022MLreview-3,freitas2022mlp} and already showcased their aptness for dislocation cores in bcc metals~\cite{GAP-Fe-Dragoni-PhysRevMaterial2018, maresca2018screw, PhysRevMaterials.4.040601NNpotential-bcc-Fe-core-and-Peierls, MishinTaNNPdislo2022, rinaldi2023non, wang2022-ML-pot-V, zhang-maresca-2023GAPs}. However, they require extensive DFT reference data~\cite{deringer2019MLreview-1,mishin2021MLreview-2,kulik2022MLreview-3,freitas2022mlp}, and magnetic ML potentials are nascent~\cite{eckhoff2021magMLp,drautz2020magace, novikov2022magMLp, chapman2022-magnetic-MLIP,rinaldi2023non, Kotykhov2023-mag-MTP-FeAl, yuan2024magnetic-NNP}.

Bond order potentials (BOPs) are interatomic potentials based on tight-binding model~\cite{DraPett06, DraPettMagBOP11}. They suffice modeling systems with tens of thousands of atoms and, in addition, treat magnetism explicitly. BOPs proved their aptness for many transition metals~\cite{MnMagBOPDrain14,Ford-14,MirMoTaNbW14,TiAlbFer19, PhysRevMaterials.8.013803-aparna-Re} and, specifically,  for $\frac{1}{2}$[111] screw dislocations in W~\cite{MroWBOP}, Mo~\cite{MroMoBOP}, and Fe~\cite{MroFePRL11}. The core structures match DFT results in every case, evincing BOPs' reliability for dislocations.

We recently developed an accurate and transferable magnetic BOP for Fe-Co alloys based on DFT reference data~\cite{EgorovBOP23}. Here, we employed it for large-scale atomistic simulations of screw dislocations in ordered ferromagnetic B2 FeCo. First, to vindicate BOP's validity for dislocations, we tested how accurately it predicts the $\gamma$~surface of the slip plane.
We then assessed the partial dislocations' equilibrium separation and elucidated their cores' symmetry. We also examined how local magnetic moments and electronic structure change in the cores.

\section{\label{sec:simuldetails}Simulation details}

\subsection{Computational details}

We used the~{\UrlFont{VASP}} package~\cite{vasp-1,vasp-2,vasp-3} for DFT calculations with the projector augmented wave (PAW) method for pseudopotentials~\cite{PAW}, PBE (Perdew-Burke-Ernzerhof) exchange-correlation functional~\cite{PBE}, 400~eV cut-off energy, and dense Monkhorst-Pack k-point meshes~\cite{Kmesh}. We used the~{\UrlFont{BOPfox}} package~\cite{bopfox} for analytic BOP calculations with the same settings as in Ref.~\cite{EgorovBOP23} and employed the~{\UrlFont{LAMMPS}}~\cite{LAMMPS} package implemented in the Atomic Simulation Environment (ASE)~\cite{larsen2017ase} for MEAM. Atomic positions were relaxed using the FIRE algorithm~\cite{FIRE_Bitzek_PhysRevLett.97.170201} until forces were less than 0.03~eV/Å for the $\gamma$ surface calculations and 0.003~eV/Å for the core structures. We used a weaker convergence criterion for the $\gamma$ surfaces to speed up DFT calculations while fast-computing BOP allowed us to use a tighter convergence criterion for the core structures.

\subsection{Setup for computing the $\gamma$ surface}

To compute the $\gamma$ surface for a slip plane,
we adopted a periodic simulation cell, cut it in half on a $\{110\}$ plane, displaced one part of the bicrystal with respect to the other half in all directions, 
and calculated the energy as a function of the displacement. The energy difference with respect to a bulk crystal, divided by the area of the cut plane, is the generalized stacking fault energy (GSFE)~\cite{Vitek1968GSFE}. In all calculations, we relaxed the positions of atoms in the direction perpendicular to the cut plane only, as relaxation in other directions would annihilate the stacking fault~\cite{Vitek1968GSFE, MroWBOP, MroMoBOP}. For the $\gamma$ surfaces, we computed energies on a 9×9 grid; the discrete values of the energies were then converted into smooth contour plots through extrapolation (Fig.~\ref{fig:gamma_surface}).

\begin{figure}[!tb]
\begin{center}
\includegraphics[width=0.47\textwidth]{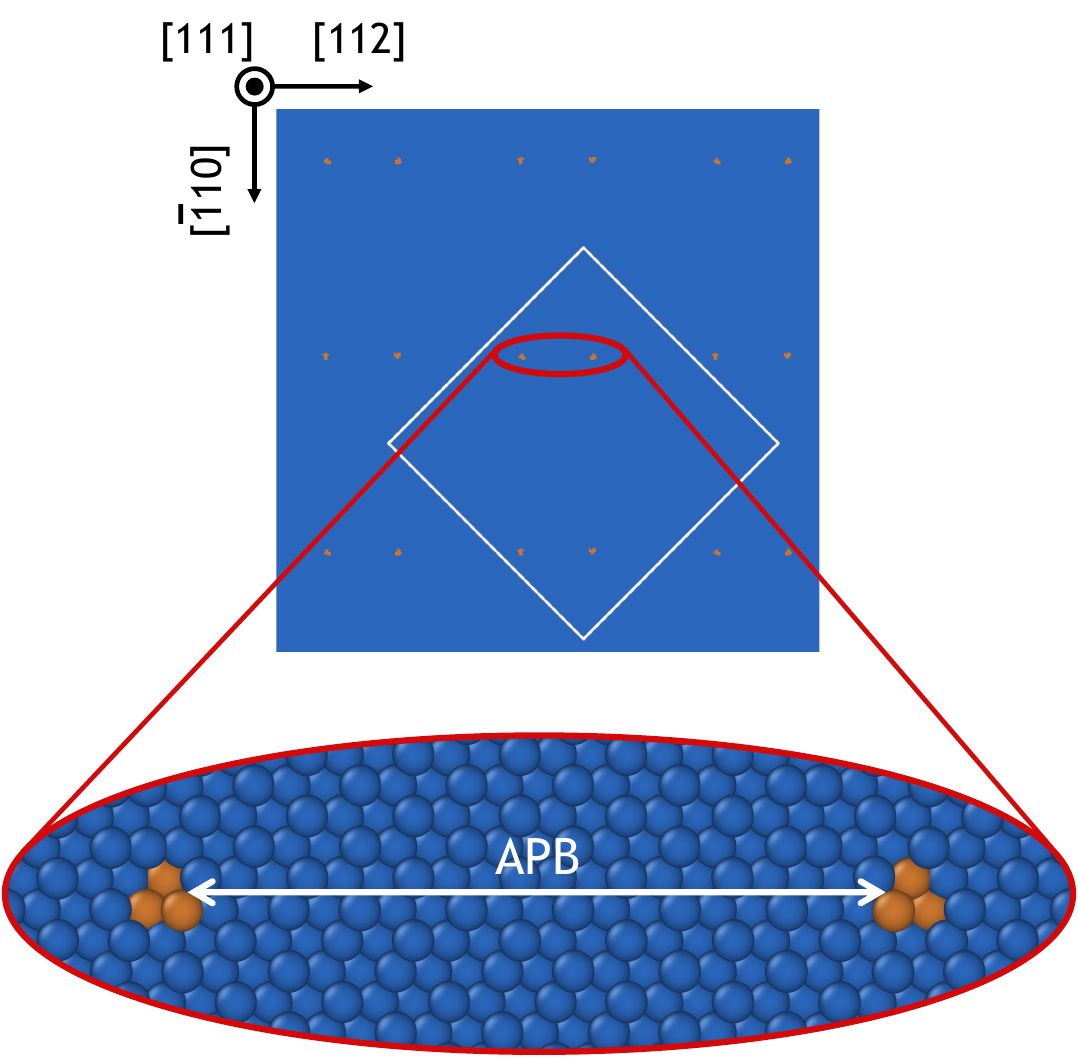}
\caption{15048-atom periodic simulation cell (white square) for the core structure of partial dislocations in B2 FeCo. Two pairs of partial $\frac{1}{2}$[111] screw dislocations in $\text{\{}110\text{\}}$ plane were introduced into the cell resulting in their quadrupolar arrangement. The blue atoms are the bulk B2, and the orange is the dislocation cores~\cite{CNA-comment}. Partials are separated by an antiphase boundary (APB) where the B2 order is distorted. The thickness of the cell is 2×$\frac{1}{2}$[111] or 4.93 Å}
\label{fig:simcell}
\end{center}
\end{figure}

\subsection{Setup for dislocation core structures}

To clarify if and how the [111]~screw superdislocations dissociate, we adopted a periodic cell containing two superdislocations with opposite Burgers vectors, resulting in their quadrupolar arrangement (see Fig.~1 in Supplemental Material~\cite{Suppl})~\cite{ventelon2007corequadr, PhysRevLett.102.055502-Clouet-2009, Clouet2020-ab-initio}. To analyze the core structure of partials, we replaced every [111] screw superdislocation with a pair of ordinary $\frac{1}{2}$[111] screw dislocations at a distance, retaining the same quadrupolar arrangement (see Fig.~\ref{fig:simcell}).
Dislocations were introduced into the simulation cell by displacing atoms according to isotropic elasticity theory using the~{\UrlFont{babel}} package~\cite{babel}. A homogeneous strain is applied to the simulation cell to level out the strain caused by the introduced dislocations~\cite{Clouet2020-ab-initio, Comptes_Rendus_Physique_2021_clouet, Kraych2019npj}.
To get the equilibrium configurations, we relaxed all atomic positions. The thickness of the simulation cell is $a$[111], equal to the Burgers vector of the superdislocation. To reduce dislocation-dislocation interactions, we tested the convergence by varying cell sizes in the directions perpendicular to the dislocation line. The resulting 132.39×132.76×4.93 Å simulation cell contained 15048 atoms (white square in Fig.~\ref{fig:simcell}). For an illustrative example of the core structure in bcc Fe (Fig.~\ref{fig:dislo_core_example}), we used a 135-atom simulation cell with a quadrupolar dislocations arrangement (see Fig.~S1a in Ref.~\cite{Kraych2019npj}).

\section{\label{sec:results}Results and discussion}

\subsection{$\gamma$ surfaces}
We first evaluated the $\gamma$~surfaces, which portray the lattice resistance against slip for a given plane. A BOP $\gamma$~surface matching the DFT would vindicate the BOP efficacy for dislocations because interatomic potentials, which produce accurate $\gamma$~surfaces, are usually also accurate for core structures~\cite{freitas2022mlp}. The DFT reference data for this BOP lacked the $\gamma$~surfaces~\cite{EgorovBOP23}; thus, verifying if the BOP could predict them accurately was crucial.

\begin{figure*}[!tb]
    \centering
    \includegraphics[width=0.97999\textwidth]{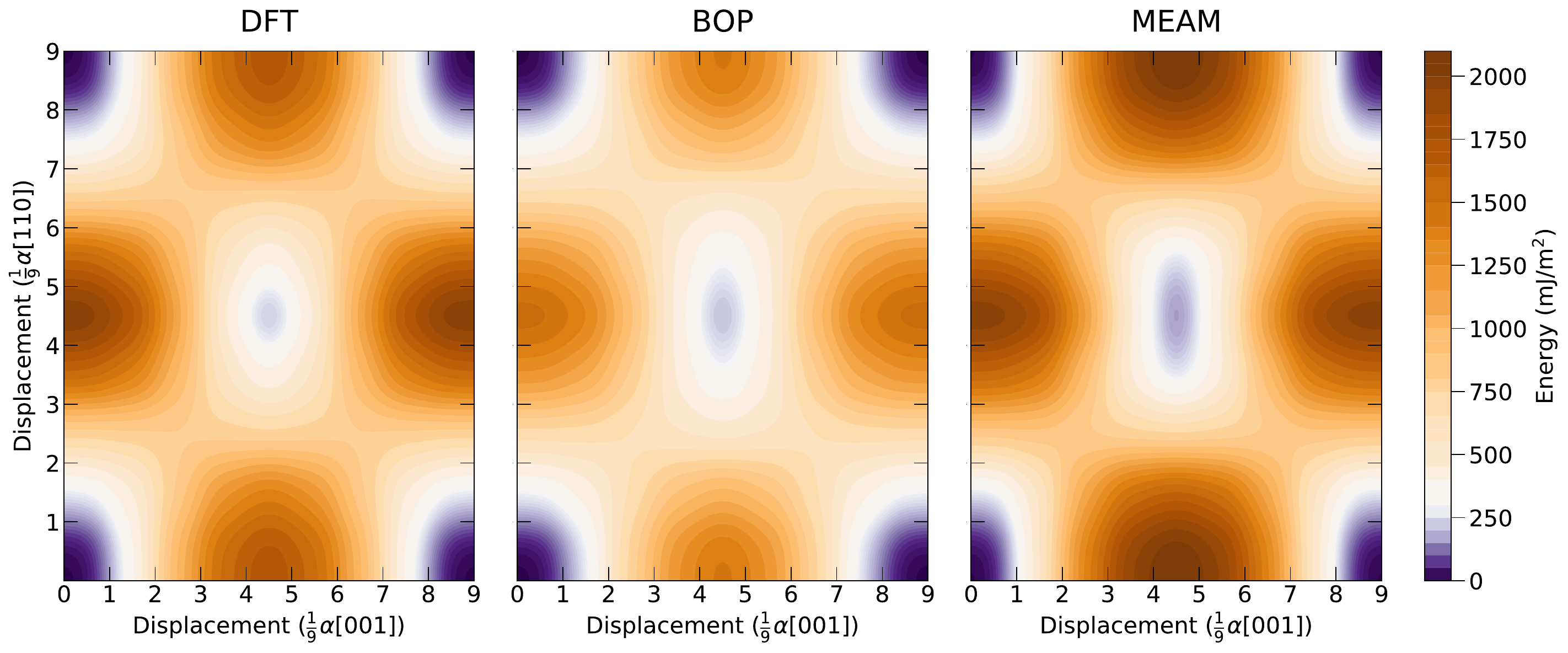}
    \caption{\label{fig:gamma_surface} Relaxed $\gamma$ surfaces for the $\text{\{}110\text{\}}$ plane (experimental slip plane~\cite{YAMAGUCHI1982}) in B2 FeCo. Diagonals on these plots are the [111] direction—experimental slip direction~\cite{YAMAGUCHI1982}. Local minima at the middle correspond to an antiphase boundary (APB) with distorted B2 order obtained when we shift two crystal parts concerning each other by $\frac{1}{2}a$[111] ($a$ is a lattice parameter).}
  \end{figure*}
We obtained the relaxed $\gamma$ surfaces for the $\{110\}$ plane—the preferable slip plane~\cite{YAMAGUCHI1982}—with BOP, DFT, and MEAM~\cite{Choi2017meam}. All three produce similar $\gamma$ surfaces (Fig.~\ref{fig:gamma_surface}), with a local minimum at the center corresponding to the antiphase boundary (APB). A single minimum indicates that the superdislocation dissociates into two partials with $\frac{1}{2}a$[111] Burgers vectors~\cite{PAIDAR_CAK}, precisely what most studies have assumed for B2 FeCo~\cite{MarcinkowskiChessin1964}.

In bcc metals and alloys, the curvature of the $\gamma$ surface contour lines hints at the symmetry of the dislocation cores. Circular contour lines between two minima along [111] direction lead to non-degenerate cores; conversely, the contour lines deviating from a circular lead to degenerate cores (see, for example, Fig.~2 in Ref.~\cite{Romaner_W-Re_PRL2010} or Fig.~1 in Ref.~\cite{DUESBERYVITEK1998CURVATURE}). For all three cases in Fig.~\ref{fig:gamma_surface}, the contour lines crossing the [111] direction deviate from a perfect circular shape, hinting at degenerate cores in B2 FeCo.

Also, obtaining an accurate equilibrium separation between partial dislocations relies heavily on the accurate APB energies (local minima at the middle), and, for BOP and DFT, they are close (see Table~\ref{tab:inputparametersandsep}). The MEAM predicts far lower APB energy, conceivably due to a lack of magnetism. 
Excess energy after the $\frac{1}{2}$[111] shift in the \text{\{}110\text{\}} planes could result from an improper B2 order of two constituent chemical species but also from the disturbed magnetic order alone, like in antiferromagnetic bcc Cr~\cite{BIENVENU2020570bccCr-1, PhysRevB.107.134105bccCr-2}. A marked increase of the APB energy due to magnetism was also observed in ordered L1$_2$ Ni$_3$Al~\cite{XU2023118986-Grabovwski-Acta-Ni3Al-APB-Magn}.
Thus, we surmise that considering magnetism in our study is vindicated.

We also computed APB energy for $\{211\}$ planes. Compared to $\{110\}$, it is about 20-30\% higher. This difference entices the partials to dissociate in the $\{110\}$ plane. For the $\{211\}$ plane, BOP again closely matches a DFT ABP energy, while MEAM yields a far lower value (see Table~\ref{tab:inputparametersandsep}). 

To summarize, the BOP predictions above align with DFT, both for the topology of the $\gamma$ surface and for the APB energies, which instills confidence that this BOP is well-suited for studying dislocations in B2 FeCo.

\subsection{\label{sec:equilsep}Equilibrium separation of the partials}

To dissect the cores of the partial dislocations in B2 FeCo, we first need to know how far apart they are. Isotropic elasticity theory can roughly estimate the equilibrium separation from the balance between partial dislocations elastic repulsion and the APB energy (see Eq.~A7 in Ref.~\cite{SeparCalcPhysRevB2013}). 
The elastic equilibrium separation (Table~\ref{tab:inputparametersandsep}) we calculated with BOP input (53~Å) is lower than with DFT input (79~Å) due to the lower shear modulus, which, along with APB energy, defines the separation (see Eq.~A7 in Ref.~\cite{SeparCalcPhysRevB2013}). Nevertheless, the BOP result is close to 45~Å calculated with experimental input (though serendipitously because BOP shear modulus and APB energy, which are proximate to experimental ones, were not in the reference data for this BOP~\cite{EgorovBOP23}, and their values are predictions).
The low APB energy obtained with MEAM yields a too large separation of 116~\AA, close to an experimental 125~\AA~for partially ordered FeCo with long-range order parameter $S$=0.59~\cite{MarcinkowskiChessin1964}. \begin{table}[!hb]
\caption{\label{tab:inputparametersandsep}
Separation of the partial $\frac{1}{2}$[111] screw dislocations in B2 FeCo, $d_{\mathrm{APB}}$, (\Rnum{1}) from isotropic elasticity theory (elastic) with input data required for calculation: equilibrium lattice parameters, $a$, Voigt-Reuss-Hill shear modulus, $G_\mathrm{{VRH}}$~\cite{Voigt-Reuss-Hill}, and antiphase boundary (APB) energy, $\gamma_\mathrm{{APB}}$, for $\{110\}$ and $\{211\}$ planes and (\Rnum{2}) separation from atomistic simulations (details on Fig.~\ref{fig:equilsep}).
}
\bgroup
\def\arraystretch{1.5}%
\begin{ruledtabular}
\begin{tabular}{cccccccc}
&$a$&$G_\mathrm{{VRH}}$&$\gamma_\mathrm{{APB_{\{110\}/\{211\}}}}$ & \multicolumn{2}{c}{$d_{\mathrm{APB}}$, (\AA)} \\
&(\AA)&(GPa)&($\mathrm{mJ/m^{2}}$) & elastic & atomistic\\
\hline
BOP\footnotemark[1] & 2.845 & 73 & 132/152 & 53 & 49  \\
DFT\footnotemark[1] & 2.844 & 93 & 114/171 & 79 & -  \\
DFT\footnotemark[2] & 2.843 & 91 & - & - & -  \\
DFT\footnotemark[3] & - & - & 129/169 & - & -  \\
MEAM\footnotemark[4] & 2.859 & 83 & 70/81 & 116 & -  \\
Exp. & 2.857\footnotemark[5] & 72\footnotemark[5] & 157\footnotemark[6]/- & 45 & -  \\
\end{tabular}
\end{ruledtabular}
\egroup
\footnotetext[1]{This work}
\footnotetext[2]{Reference \cite{Jain2013materialsproject}}
\footnotetext[3]{Reference \cite{KrchmarPRB06}}
\footnotetext[4]{Reference \cite{Choi2017meam}}
\footnotetext[5]{Reference \cite{belousov2009expB}, measured at 293 K}
\footnotetext[6]{Reference \cite{marcinkowski1963ExpAPBener}}
\end{table}

\begin{figure}[!hb]
\begin{center}
\includegraphics[width=0.479\textwidth]{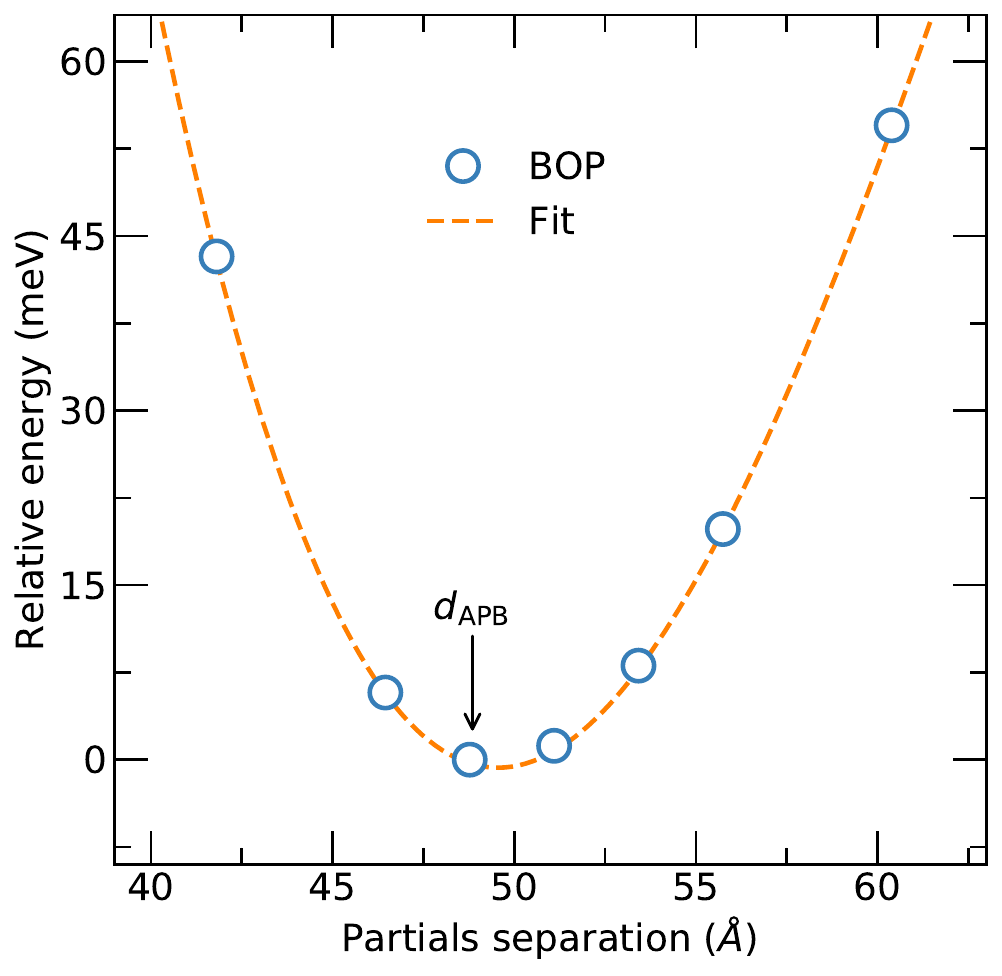}
\caption{Relative energy of the simulation cell versus separation of the partial dislocations obtained using BOP. The minimum corresponds to the equilibrium separation of the partial dislocations, $d_{\mathrm{APB}}$ (see also Table~\ref{tab:inputparametersandsep}).}
\label{fig:equilsep}
\end{center}
\end{figure} In the atomistic simulations with BOP (at 0K), [111] superdislocations do not dissociate unless the atomic positions are randomly distorted (Then, BOP correctly predicted that during relaxation, every [111] superdislocation dissociates into two $\frac{1}{2}$[111] partials connected by APB. Random distortions led partials astray, and they could move both on $\{110\}$ and $\{211\}$ planes,  
depending on the distortion.
See more details in Supplemental Material~\cite{Suppl}). Undistorted superdislocation stays undissociated because, at 0 K, partials cannot surmount a Peierls barrier. Therefore, we varied distances between partials on $\{110\}$ plane and, again, after relaxation, evaluated the cell's relative energies (Fig.~\ref{fig:equilsep}).
The curve attains equilibrium separation at 49~Å, with energy much lower than that of undissociated superdislocation (or any randomly distorted dissociated configuration), manifesting that it is the ground state. Besides, the equilibrium separation agrees closely with the elasticity theory (cf.~Table~\ref{tab:inputparametersandsep}). The slight difference between the elastic and atomistic separations may be attributed to the finite size of the cores, atomic interactions, or anisotropic effects.

\begin{figure*}[!htb]
    \centering
    \includegraphics[width=0.999999\textwidth]{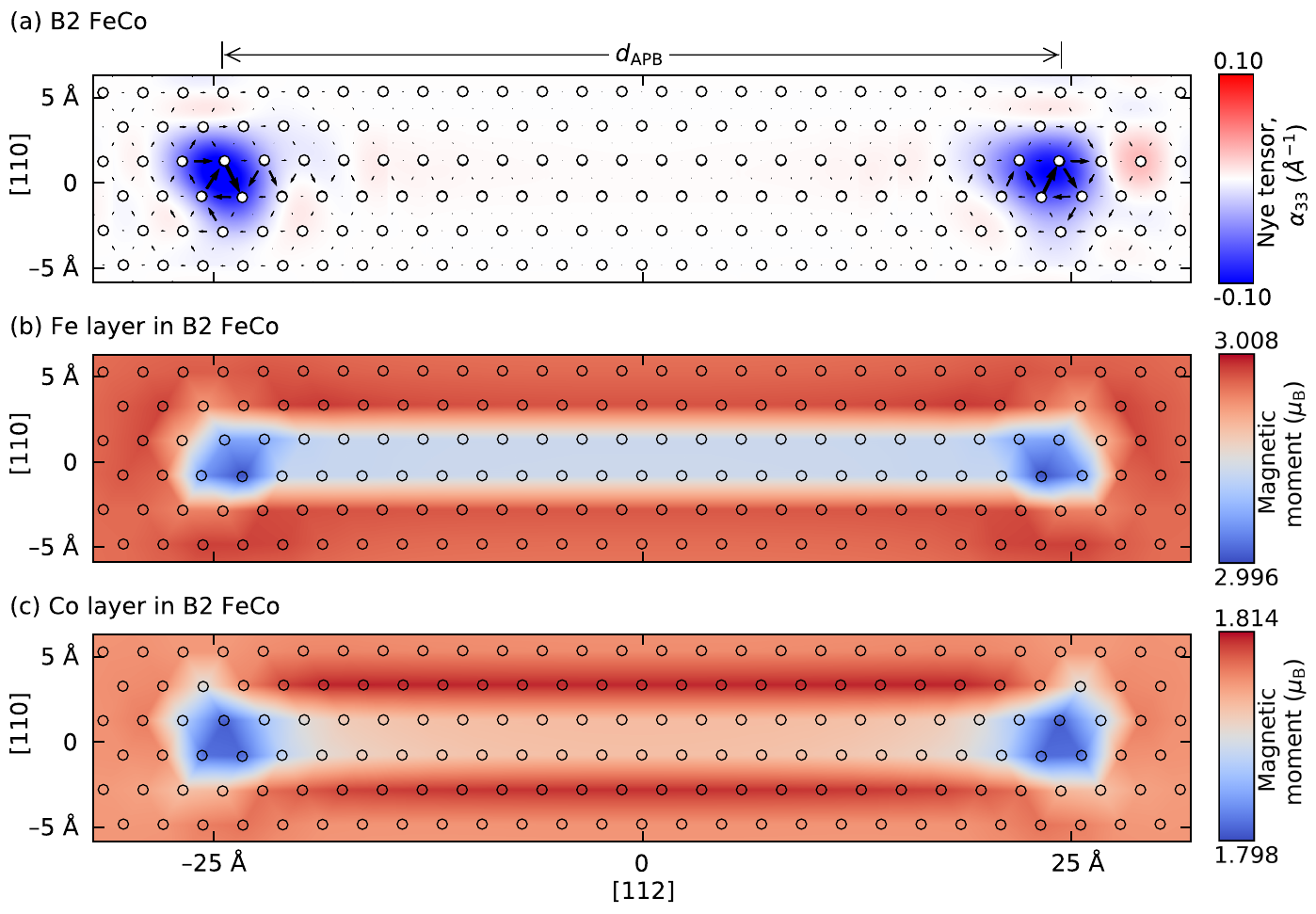}
    \caption{\label{fig:core_struc} The core structure of partial $\frac{1}{2}$[111] screw dislocations in B2 FeCo we obtained with BOP~\cite{visual-comment}. Partials separated by an antiphase boundary (APB) of the length $d_{\mathrm{APB}}$ = 49~Å (See Section~\ref{sec:equilsep}). The differential displacement and Nye tensor distribution (a) reveal degenerate core structures~\cite{red-dot-comment}. Additionally, the local magnetic moments of Fe and Co layers in B2 FeCo [(b) and (c)] display lower magnetic moments within the cores and APB compared to bulk.}
\end{figure*}

\subsection{Dislocation core structures}

Employing the equilibrium separation of the partial dislocations ($d_{\mathrm{APB}}$ in Fig.~\ref{fig:equilsep}), we obtained their relaxed cores with BOP. We then scrutinized the core structures with a differential displacement map~\cite{DDPVitek1970} and the screw component ($\alpha_{33}$) of the Nye tensor~\cite{NYE1953153}, which portrays the continuous distribution of the Burgers vector~\cite{hartley2005.Nye-tensor-1, hartley2005-Nye-tensor-2, mendis2006-core-experiment-2-and-Nye-tensor}. We revealed that both cores attain distorted degenerate structures spread predominantly along the APB fault [Fig.~\ref{fig:core_struc}(a)]. Romaner $et$~$al.$ with DFT obtained similar cores for individual $\frac{1}{2}$[111] screw dislocations in disordered bcc Fe-Co~\cite{Romaner_FeCo_2014}. The authors also observed a change from the non-degenerate core at low Co concentrations, starting from zero, that is, from pure Fe to the degenerate core at around 1:1 composition (worth noting that for pure Fe, our BOP also correctly predicts the non-degenerate core; see Fig.~\ref{fig:dislo_core_example}). 

Core symmetry in bcc-like Fe-Co alloys may differ from the one in pure Fe, with its non-degenerate core, due to pure Co being hcp. Bcc alloys, where $both$ constituents reside in a bcc ground state, such as W-Mo or W-Ta, retain non-degenerate cores~\cite{Alling2020MoWcore, RomanerActaMatW-Tacore2012}. However, if one constituent resides in another ground state (as in W-Re, where pure Re is hcp), the degenerate core structure seems preferable~\cite{Romaner_W-Re_PRL2010,samolyuk2012cores-alloys}. We confirmed the same here.

Additionally, in a recent work, Wang $et$~$al.$ linked the core structure in bcc metals and alloys to the energy differences between their bcc and fcc phases, expressed in the materials index $\chi$~\cite{wang2022taming}. $\chi$ emanates from the bcc-fcc energy difference in the pure bcc metal (in our case, Fe) and the same in the alloy (in our case, the energy difference between bcc-based B2 and fcc-based L1$_0$ phases in FeCo). These differences computed with BOP closely match the DFT values~\cite{EgorovBOP23}. A resulting $\chi$ index of roughly 0.69 corresponds to the degenerate core structure (near the transition from non-degenerate)~\cite{wang2022taming}, which is what we observed in B2 FeCo.

\begin{figure*}[!tb]
\centering
\begin{center}
\includegraphics[width=0.8\textwidth]{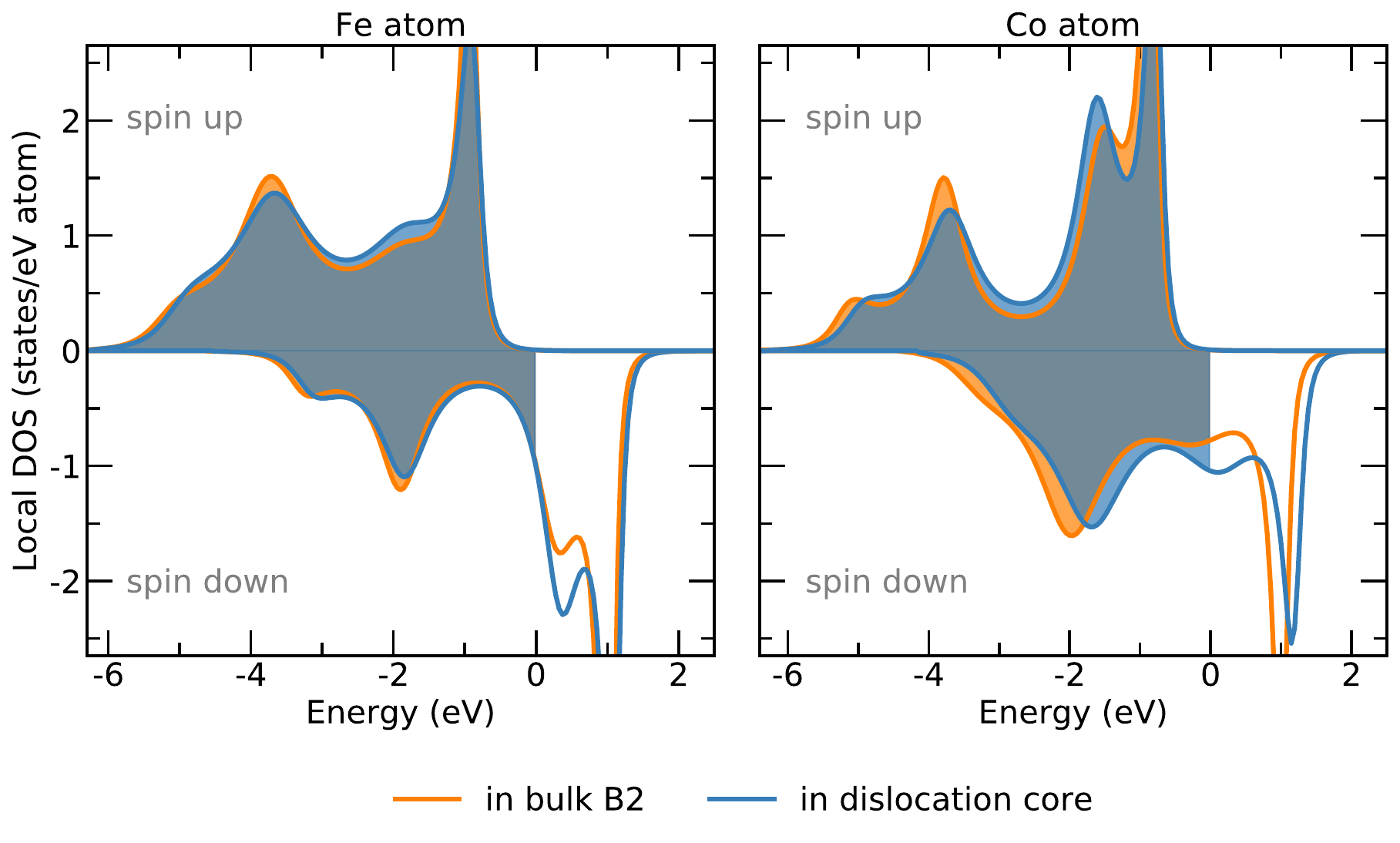}
\caption{Local density of states (DOS) of Fe and Co atoms in the dislocation core and bulk B2 FeCo computed with BOP.
}
\label{fig:localdos}
\end{center}
\end{figure*}

As magnetism defines the Fe-Co phase stability~\cite{AbrikFeCoPRB, FahnleAntisite}, we also looked at how the local magnetic moments change in the dislocation cores.
In disordered bcc Fe-Co, as Romaner $et$~$al.$ observed with DFT, they decrease~\cite{Romaner_FeCo_2014}. The authors linked it to the $d$-band filling, and we can expect the same due to a similar band filling in B2 FeCo~\cite{PhysRevB.83.054202Rahaman}. Indeed, we observed that the magnetic moments of both Fe and Co atoms in the core decreased [Fig.~\ref{fig:core_struc}(b-c)]—moreover, the decrease within 1\% correlates with the decrease in bcc Fe-Co~\cite{Romaner_FeCo_2014}.  
(Tight-binding based methods, such as BOP, provide reliable energy differences between competing magnetic and non-magnetic phases, but the exact values of magnetic moments are not always robust~\cite{PhysRevB.71.174115-TB-local-magmoms-and-phase-stab}. Thus, we should approach magnetic moment predictions with a grain of salt; even DFT calculations with different exchange-correlation functionals render different results for magnetic moments in the core~\cite{VENTELON20133973-bcc-Fe-magmoms-dft, Fe-Magmoms-in-core-JApPhy, Romaner_FeCo_2014, PhysRevB.102.094420-alling-dislo-bcc-fe-dft, Frederiksen_cores_DFT_2003}.)

We also examined how dislocations modify the electronic structure, namely, the local density of states (DOS) of the atoms in the cores. For the bulk B2, BOP predicts magnetic and non-magnetic DOS, which is consistent with DFT (see Fig.~3 in Ref.~\cite{EgorovBOP23}); thus, we can rely on the BOP predictions. For the local DOS in the dislocation core, we observed moderate changes for the Fe atoms and distinct for the Co atoms (Fig.~\ref{fig:localdos}). For Co, some lower energy states (for spin down) shift closer to the Fermi level compared to the bulk B2, thus increasing the band energy. As Dezerald~$et~al.$ unraveled, such an increase affects the core energies and the Peierls energies of dislocations in bcc transition metals, increasing them too~\cite{Dezerald-dos-in-core}. We can expect similar dependency in transition metal alloys, including B2 FeCo. (We will present the Peierls energy for B2 FeCo and compare it with those in disordered FeCo and pure Fe in a separate publication.)

\section{\label{sec:discussion}Summary and conclusions}

Using a magnetic bond-order potential (BOP), we determined the atomic core structure of $\frac{1}{2}$[111] screw dislocations in ordered B2 FeCo and can draw the following main conclusions: 

(\Rnum{1}) The $\gamma$ surface for the $\{110\}$ slip plane obtained using BOP is consistent with DFT.

(\Rnum{2}) Screw dislocations in B2 FeCo exist in pairs, separated by a 50~Å wide antiphase boundary. This large separation obstructs DFT simulations while it is reachable for BOP.

(\Rnum{3}) $\frac{1}{2}$[111] screw dislocations in B2 FeCo—unlike most bcc transition metals but likewise disordered FeCo~\cite{Romaner_FeCo_2014}—accommodate degenerate core structures.

(\Rnum{4}) Magnetic moments decrease in the cores; just as in disordered FeCo alloys~\cite{Romaner_FeCo_2014}.

(\Rnum{5}) Dislocations alter the local DOS—and thus atomic interactions—in the cores.

(\Rnum{6})  Significant alterations in the local DOS for Co atoms in the core (unlike Fe atoms) are expected to increase band energy and, hence, the Peierls energy.

\section*{Acknowledgment}
We thank Lorenzo La Rosa and Francesco Maresca for their valuable comments on the manuscript. We acknowledge financial support from the International Max-Planck Research School SurMat and the Wilhelm and Günter Esser Foundation. The DFG supported part of this work within the DFG-ANR project MAGIKID (No. 316673557).

\bibliography{ref}

\begin{thebibliography}{126}
\bibitem [{vis()}]{visual-comment}%
  \BibitemOpen
  \href@noop {} {}\bibinfo {note} {Differential displacement, {Nye} tensor, and magnetic moments are visualized with the~{\UrlFont{atomman}} package~\cite{atomman}}\BibitemShut {NoStop}%
\bibitem [{red()}]{red-dot-comment}%
  \BibitemOpen
  \href@noop {} {}\bibinfo {note} {The pale red dot at the right edge results from numerical extrapolation to create a smooth plot from discrete values of the Nye tensor and does not impart significant meaning}\BibitemShut {NoStop}%
\end{thebibliography}

\end{document}